\RequirePackage[2020-02-02]{latexrelease}
\documentclass{aa}  

\usepackage[switch]{lineno}
\usepackage{natbib}
\bibpunct{(}{)}{;}{a}{,}{,}
\usepackage{longtable}

\usepackage{graphicx,subcaption}
\usepackage{pifont,txfonts}

\usepackage{aas_macros}

\usepackage{xspace}
\usepackage{pstricks}
\usepackage{color}
\graphicspath{{figures/}}
\usepackage{pgf}

\usepackage{hyperref}
\hypersetup{
    unicode=true,           
    pdftoolbar=true,        
    pdfmenubar=true,        
    pdffitwindow=true,      
    pdftitle={Dumbbell (17365) Thymbraeus},
    pdfauthor={},
    pdfsubject={Planetary Science},
    pdfkeywords={},         
    pdfnewwindow=true,      
    colorlinks=true,        
    linkcolor=gray,         
    citecolor=blue,         
    filecolor=gray,         
    urlcolor=gray           
} 

\newcommand{\numb}[1]{\textcolor{orange}{#1}}
\renewcommand{\numb}[1]{#1}  

\newcommand{\textsub}[1]{\ensuremath{_{\textrm{#1}}}}

\newcommand{\bsep}{\ensuremath{\beta_{\textrm{SEP}}}}

\newcommand{\ssodnet}{\texttt{SsODNet}\xspace}

\newcommand{\Dens}{\numb{830}}
\newcommand{\dDens}{\numb{50}}
\newcommand{\sid}{kg.m$^{-3}$}

\newcommand{\revadd}[1]{\textbf{#1}}
\newcommand{\revrm}[1]{}
\renewcommand{\revadd}[1]{#1}

\usepackage{catoptions}
\makeatletter

\def\Autoref#1{%
  \begingroup
  \edef\reserved@a{\cpttrimspaces{#1}}%
  \ifcsndefTF{r@#1}{%
    \xaftercsname{\expandafter\testreftype\@fourthoffive}
      {r@\reserved@a}.\\{#1}%
  }{%
    \ref{#1}%
  }%
  \endgroup
}
\def\testreftype#1.#2\\#3{%
  \ifcsndefTF{#1autorefname}{%
    \def\reserved@a##1##2\@nil{%
      \uppercase{\def\ref@name{##1}}%
      \csn@edef{#1autorefname}{\ref@name##2}%
      \autoref{#3}%
    }%
    \reserved@a#1\@nil
  }{%
    \autoref{#3}%
  }%
}
\makeatother

\begin{document} 
    \title{Close-to-fission dumbbell Jupiter-Trojan (17365) Thymbraeus%
    }
    
  \author{%
    B.~Carry\inst{1}               \and 
    P.~Descamps\inst{2}          \and 
    M.~Ferrais\inst{3}              \and 
    J.-P.~Rivet\inst{1}            \and 
    J.~Berthier\inst{2}          \and 
    E.~Jehin\inst{4}             \and 
    D.~Vernet\inst{5}          \and 
    L.~Abe\inst{1}                 \and 
    P.~Bendjoya\inst{1}            \and 
    F.~Vachier\inst{2}           \and 
    M.~Pajuelo\inst{2,6} \and 
    M.~Birlan\inst{2,7} \and 
    F.~Colas\inst{2}             \and 
    Z.~Benkhaldoun\inst{8}             
  }


   \institute{
     Universit\'e C{\^o}te d'Azur, Observatoire de la
     C{\^o}te d'Azur, CNRS, Laboratoire Lagrange, France
     \email{benoit.carry@oca.eu}\label{oca}
     \and 
     IMCCE, Observatoire de Paris, PSL Research University, CNRS, Sorbonne Universit{\'e}s, UPMC Univ Paris 06, Univ. Lille, France
     \label{imcce}
     \and 
     Arecibo Observatory, University of Central Florida, HC-3 Box 53995, Arecibo, PR 00612, USA
     \label{ao}
     \and 
     Space sciences, Technologies and Astrophysics Research (STAR) Institute, Universit{\'e} de Li{\`e}ge, All{\'e}e du 6 Ao{\^u}t 17, 4000 Li{\`e}ge, Belgium
     \label{liege}
     \and 
     Universit\'e C{\^o}te d'Azur, Observatoire de la
     C{\^o}te d'Azur, CNRS, UAR Galilée, France
     \label{galilee}
     \and 
     Secci{\'o}n F{\'i}sica, Departamento de Ciencias, Pontificia Universidad Cat{\'o}lica del Per{\'u}, Apartado 1761, Lima, Per{\'u}
     \label{puc}
     \and 
     Astronomical Institute of the Romanian Academy, Cutitul de argint -5 , sector 4, Bucharest, Romania
     \label{buca}
     \and 
     Oukaimeden Observatory, High Energy Physics and Astrophysics Laboratory, Cadi Ayyad University, Marrakech, Morocco
     \label{ouka}
  }
  \date{Received X; accepted Y}

  \abstract
   {Every population of small bodies in the Solar system contains a 
   sizable fraction of multiple systems. Among these, the Jupiter Trojans
   have the lowest number of known binary systems and the least characterized.}
   {We aim at characterizing the reported binary system
    (\revadd{17365}) Thymbraeus, one of the \revadd{only seven} multiple systems known 
    among Jupiter Trojans.}
   {We conducted light curves observing campaigns in 2013, 2015, and 2021 
   with ground-based telescopes. We model these lightcurves using
   dumbbell equilibrium figures.}
   {We show that Thymbraeus is unlikely a binary system. Its light curves
   are fully consistent with a \revadd{bilobated shape: a} dumbbell equilibrium figure.
   We determine a low density of $\Dens \pm \dDens$\,\sid{}, consistent
   with the reported density of other Jupiter Trojan asteroids and
   small Kuiper-belt objects.
   The angular velocity of Thymbraeus is close to fission. If separated,
   its components would become a similarly-sized double asteroid such 
   as the other Jupiter Trojan (617) Patroclus.
  }
   {}


\authorrunning{Carry et al.}
\titlerunning{Dumbbell (17365) Thymbareus}
\maketitle


%
\section{Introduction}

\begin{figure}[t]
    \centering
    \input{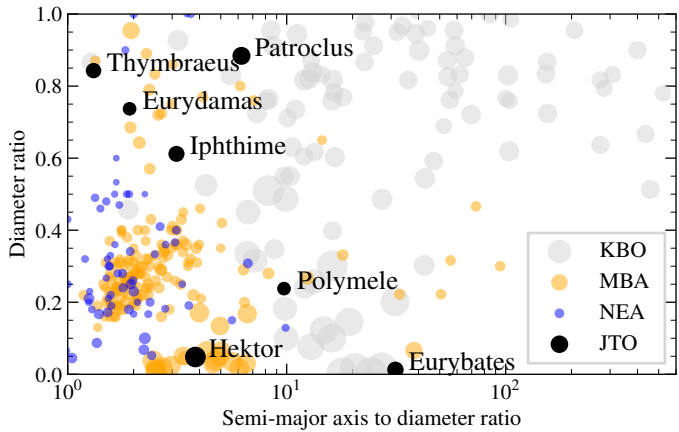}
    \caption{The diversity among small body binaries, with the
    \revadd{seven} known Jupiter 
    Trojans (JTO) in black.
    \revadd{The satellite diameters and semi-major axes are from  
    \citet{2018PDSS..305.....J}
    and the diameters of the primary bodies from
    the \ssodnet service
    \citep{2023A&A...671A.151B}.}
    }
    \label{fig:binaries}
\end{figure}

  The small bodies with satellites represent a highly diverse 
  population in the Solar system, spanning a wide range
  of diameter, separation, and size ratio
  \citep[\Autoref{fig:binaries}, and][for a review]{2015-AsteroidsIV-Margot}.
  Some systems are made of large and similarly-sized bodies.
  These double systems are thought to be primordial
  and are abundant in the Kuiper belt
  \citep{2017NatAs...1E..88F}.
  The largest small bodies (diameters above a 100\,km typically) can
  also have small satellites, believed to form from the re-accumulation
  of ejecta after impacts, found in both the asteroid and
  the Kuiper belt {\citep[e.g.,][]{
   2009AJ....137.4766R, 
   2014Icar..239..118B,
   2019A&A...623A.132C, 
   2021A&A...650A.129C,
   2022Icar..38215013V}}.
  An abundant population \citep[about 15\%,][]{2002Sci...296.1445M, 2006Icar..181...63P}
  of small asteroids (diameter below 10\,km) have close-in satellites,
  likely produced by fission due to YORP spin-up
  \citep{
    2008Natur.454..188W,
    2015-AsteroidsIV-Walsh, 
    2022NatCo..13.4589Z}.

  As of today, the least characterized population of small bodies 
  in term of multiplicity are the Jupiter Trojans.
  Only sddrev{Seven} multiple systems have been discovered:
    (617) Patroclus from Gemini \citep{2001IAUC.7741....2M}, 
    (624) Hektor from W. M. Keck \citep{2006IAUC.8732....1M}, 
    (3548) Eurybates with the Hubble Space Telescope \citep[HST,][]{2020PSJ.....1...44N}, 
    \revadd{(15094) Polymele by stellar occultation \citep{2022DPS....5451203B},} 
    (16974) Iphtime with the HST \citep{2016LPI....47.2632N},
    and both
    (17365) Thymbraeus and 
    (29314) Eurydamas from light curves \citep{2007AJ....134.1133M}.
  This low number of binary systems is most-likely the result of
  observing biases.
  Radar observations efficient in discovering satellites
  are limited in range \citep{2015aste.book..165B},
  adaptive-optics observations require a bright source and
  have been mainly limited to large main-belt asteroids
  \citep[and the brightest KBOs,][]{1999Natur.401..565M,
    2005Natur.436..822M,
    2011A&A...534A.115C,
    2020A&A...641A..80Y}.
  While the HST does not require a bright source, most studies focused
  on KBOs 
  \citep{2004Icar..172..402N, 2006ApJ...639L..43B, 2011Icar..213..678G}
  until the selection of the Lucy mission by NASA
  \citep{2017LPI....48.2025L}.
  Finally, while the majority of binaries have been discovered by light curves
  \citep[see][]{2018PDSS..305.....J},
  often by amateur astronomers, the Jupiter Trojans are faint for
  most amateur equipment
  \citep{2014ExA....38...91M}.

  However, Jupiter Trojans are a unique population, related to the outer
  Solar system, and trapped on the L4/L5 Lagragian points of the
  Sun-Jupiter system during the phase of dynamical instability
  in the early Solar system
  \citep{2005Natur.435..462M, 2018NatAs...2..878N}.
  We focus here on the reported binary
  (17365) Thymbraeus\footnote{Formerly 1978 VF11}.
  We conducted an observing campaign spanning several oppositions
  to determine the physical properties of this object.

  The article is organized as following.
  In \Autoref{sec:obs} we present the \revadd{observations} and data reduction.
  We describe how we determine the properties of Thymbraeus in 
  \Autoref{sec:dyn}, and discuss their implications in 
  \Autoref{sec:disc}.

\section{Observations\label{sec:obs}}

  \indent We observed Thymbraeus in 2015 on \numb{seven} dates with the
  104\,cm Omicron telescope of the
  Centre P\'edagogique Plan\`ete et Univers (C2PU) facility 
  \citep{2012sf2a.conf..643B} located at the Calern observing site
  (Côte d'Azur Observatory, France, IAU code: 010).
  In 2021, we used three facilities.
  We collected \numb{29} epochs with
  the 60\,cm André Peyrot telescope mounted at 
  Les Makes observatory (IAU code 181) on La Réunion Island.
  We finally acquired \numb{15 and 5} epochs with the TRAPPIST
  south and north telescopes \citep[IAU code I40 and Z53,
    respectively,][]{2011Msngr.145....2J}.
  For all these observations, standard reduction (dark subtraction,
  flat-fielding) and photometry procedure (plate solution, zero point,
  aperture photometry) was conducted.
  We also retrieved the four light curves
  obtained in 2013 by \citet{2014MPBu...41...95S} with a 40\,cm telescope
  in his observatory (IAU code U81), available
  on ALCDEF\footnote{\url{https://alcdef.org/}}
  \citep{2016MPBu...43...26W}.
  The detailed logs of observations are provided in 
  \Autoref{tab:obs}.
  \revadd{Photometric uncertainties are estimated to 
  about 0.03\,mag from their scatter.}

  We complement this data set with
  the observations from 2005 and 2006
  reported by \citet{2007AJ....134.1133M} that we digitized.
  These last two light curves were only plotted in the article
  and not available in tabular format. Furthermore, these were
  reported as function of the rotation phase and in reduced
  magnitude, not as observed (epoch, magnitude). 
  We thus do not use these two light curves for modeling but
  as a posteriori  validation of the solution.

\begin{table}
  \caption[]{Log of observations.\label{tab:obs}}
  \begin{tabular}{ccrrrr}
    \hline\hline
    Date & Telescope & Duration & N & V & $\alpha$ \\
    &&&& (mag) & (\degr) \\
    \hline
2013-09-28 & U81 &  6h52 &  69 & 17.7 &    8.0 \\
2013-09-29 & U81 &  6h26 &  48 & 17.7 &    7.9 \\
2013-10-02 & U81 &  6h17 &  63 & 17.6 &    7.4 \\
2013-10-03 & U81 &  7h16 &  55 & 17.6 &    7.2 \\
2015-11-23 & 010 &  2h00 &  24 & 18.2 &    9.0 \\
2015-11-25 & 010 &  6h20 &  74 & 18.2 &    8.7 \\
2015-12-16 & 010 &  3h15 &  39 & 17.9 &    5.7 \\
2015-12-16 & 010 &  3h25 &  41 & 17.9 &    5.7 \\
2015-12-19 & 010 &  4h25 &  53 & 17.9 &    5.2 \\
2015-12-20 & 010 &  1h55 &  23 & 17.9 &    5.0 \\
2015-12-21 & 010 &  7h50 &  94 & 17.9 &    4.8 \\
2021-05-14 & Z53 &  1h27 &  43 & 17.9 &    4.6 \\
2021-05-15 & I40 &  4h22 &  91 & 17.9 &    4.4 \\
2021-05-17 & I40 &  8h26 &  97 & 17.9 &    4.0 \\
2021-05-17 & 181 &  6h42 &  46 & 17.9 &    4.0 \\
2021-05-18 & 181 &  3h47 &  39 & 17.9 &    3.8 \\
2021-05-22 & 181 &  1h48 &   6 & 17.8 &    3.1 \\
2021-05-23 & I40 &  3h40 &  46 & 17.8 &    2.9 \\
2021-05-24 & I40 &  3h31 &  41 & 17.8 &    2.7 \\
2021-05-24 & 181 &  2h06 &  22 & 17.8 &    2.7 \\
2021-05-25 & I40 &  2h45 &  34 & 17.8 &    2.5 \\
2021-05-29 & 181 &  4h05 &  31 & 17.7 &    1.7 \\
2021-05-30 & 181 &  2h42 &  28 & 17.7 &    1.5 \\
2021-05-30 & I40 &  3h56 &  43 & 17.7 &    1.5 \\
2021-05-31 & 181 &  4h24 &  38 & 17.7 &    1.3 \\
2021-06-02 & 181 &  3h30 &  38 & 17.6 &    0.9 \\
2021-06-03 & 181 &  3h30 &  35 & 17.6 &    0.7 \\
2021-06-06 & 181 &  2h47 &  16 & 17.5 &    0.2 \\
2021-06-07 & 181 &  1h36 &  14 & 17.5 &    0.2 \\
2021-06-07 & I40 &  3h35 &  52 & 17.5 &    0.2 \\
2021-06-07 & I40 &  7h41 &  88 & 17.5 &    0.2 \\
2021-06-08 & 181 &  5h00 &  27 & 17.6 &    0.4 \\
2021-06-09 & 181 &  2h17 &  17 & 17.6 &    0.5 \\
2021-06-09 & Z53 &  2h04 &  28 & 17.6 &    0.5 \\
2021-06-10 & 181 &  5h42 &  20 & 17.6 &    0.7 \\
2021-06-12 & 181 &  7h24 &  43 & 17.7 &    1.1 \\
2021-06-13 & 181 &  6h54 &  41 & 17.7 &    1.3 \\
2021-06-14 & 181 &  4h24 &  17 & 17.7 &    1.5 \\
2021-06-15 & 181 &  3h42 &  17 & 17.7 &    1.7 \\
2021-06-17 & 181 &  6h17 &  33 & 17.7 &    2.1 \\
2021-06-19 & 181 &  3h06 &  19 & 17.8 &    2.5 \\
2021-06-26 & 181 &  0h54 &  11 & 17.9 &    3.9 \\
2021-06-27 & 181 &  5h30 &  37 & 17.9 &    4.1 \\
2021-06-29 & 181 &  5h47 &  40 & 17.9 &    4.4 \\
2021-06-29 & I40 &  8h18 & 109 & 17.9 &    4.4 \\
2021-06-29 & Z53 &  1h48 &  31 & 17.9 &    4.4 \\
2021-07-01 & I40 &  4h54 &  69 & 17.9 &    4.8 \\
2021-07-02 & 181 &  3h42 &  23 & 17.9 &    5.0 \\
2021-07-04 & Z53 &  2h17 &  40 & 18.0 &    5.3 \\
2021-07-04 & 181 &  4h17 &  26 & 18.0 &    5.3 \\
2021-07-08 & Z53 &  2h15 &  32 & 18.0 &    6.0 \\
2021-07-25 & 181 &  3h06 &  30 & 18.2 &    8.4 \\
2021-07-26 & 181 &  4h05 &  31 & 18.2 &    8.6 \\
2021-07-30 & 181 &  1h00 &   9 & 18.3 &    9.0 \\
2021-08-02 & I40 &  4h00 &  57 & 18.3 &    9.3 \\
2021-08-03 & 181 &  0h54 &   9 & 18.3 &    9.4 \\
2021-08-03 & I40 &  4h54 &  70 & 18.3 &    9.4 \\
2021-08-06 & I40 &  3h39 &  53 & 18.3 &    9.7 \\
2021-08-12 & I40 &  5h15 &  75 & 18.4 &   10.1 \\
2021-08-27 & I40 &  4h42 &  68 & 18.5 &   10.8 \\
    \hline 
  \end{tabular}
\end{table}

\section{Analysis\label{sec:dyn}}

\subsection{Synodic period}

  We use the Phase Dispersion Minimization (PDM) technique 
  \citep{1978ApJ...224..953S} to search for
  the synodic rotation period within the photometric data
  \revadd{(all epochs are lighttime corrected)}.
  We assume here that two maxima and minima occur per rotation.
  Based on a trial period, PDM bins data
  according to the rotational phase.
  The average variance of these subsets is compared
  to the overall variance of the full set of
  observations. It defines the statistical parameter
  $\theta$.
  The best estimate of the period is the one which
  minimises $\theta$. This
  method does not assume any sinusoidal variation 
  of the light curve and is well suited for unevenly
  spaced observations. 
  PDM finds all periodic components or subharmonics (aliases of 
  the period).
  \revadd{We follow the approach used by 
  \citet{2020Icar..35213990B} for (617) Patroclus:
  we first determine the synodic period with PDM for each
  epoch of observation: 2013, 2015 and 2021.
  We then search for the
  fundamental synodic period by combining all epochs,
  and find
  P\textsub{syn} = 12.671575 $\pm$ 0.000003 h.}
  %

\subsection{Spin-vector coordinates}

\begin{figure*}[t]
  \centering
  \includegraphics[width=0.5\hsize]{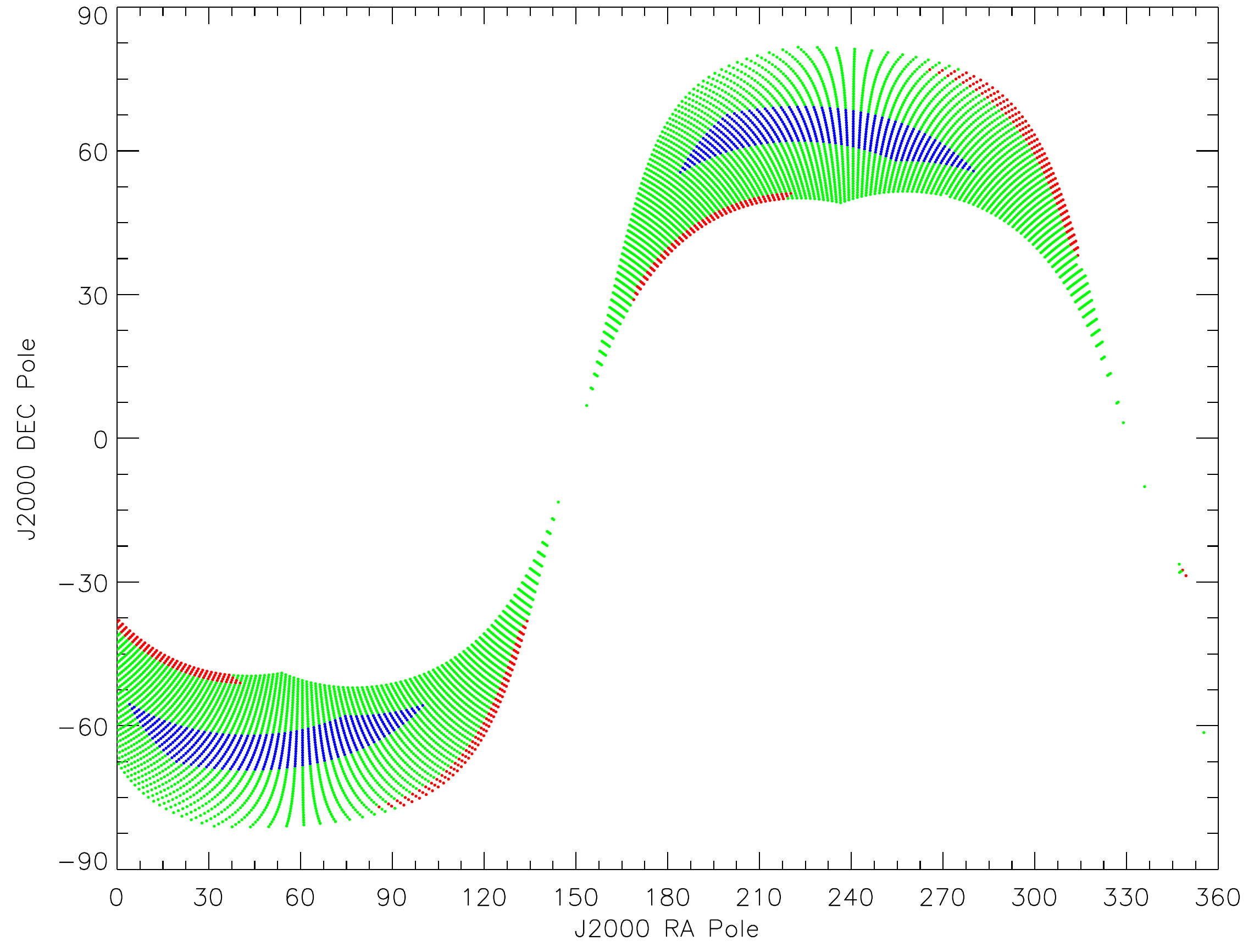}%
  \includegraphics[width=0.5\hsize]{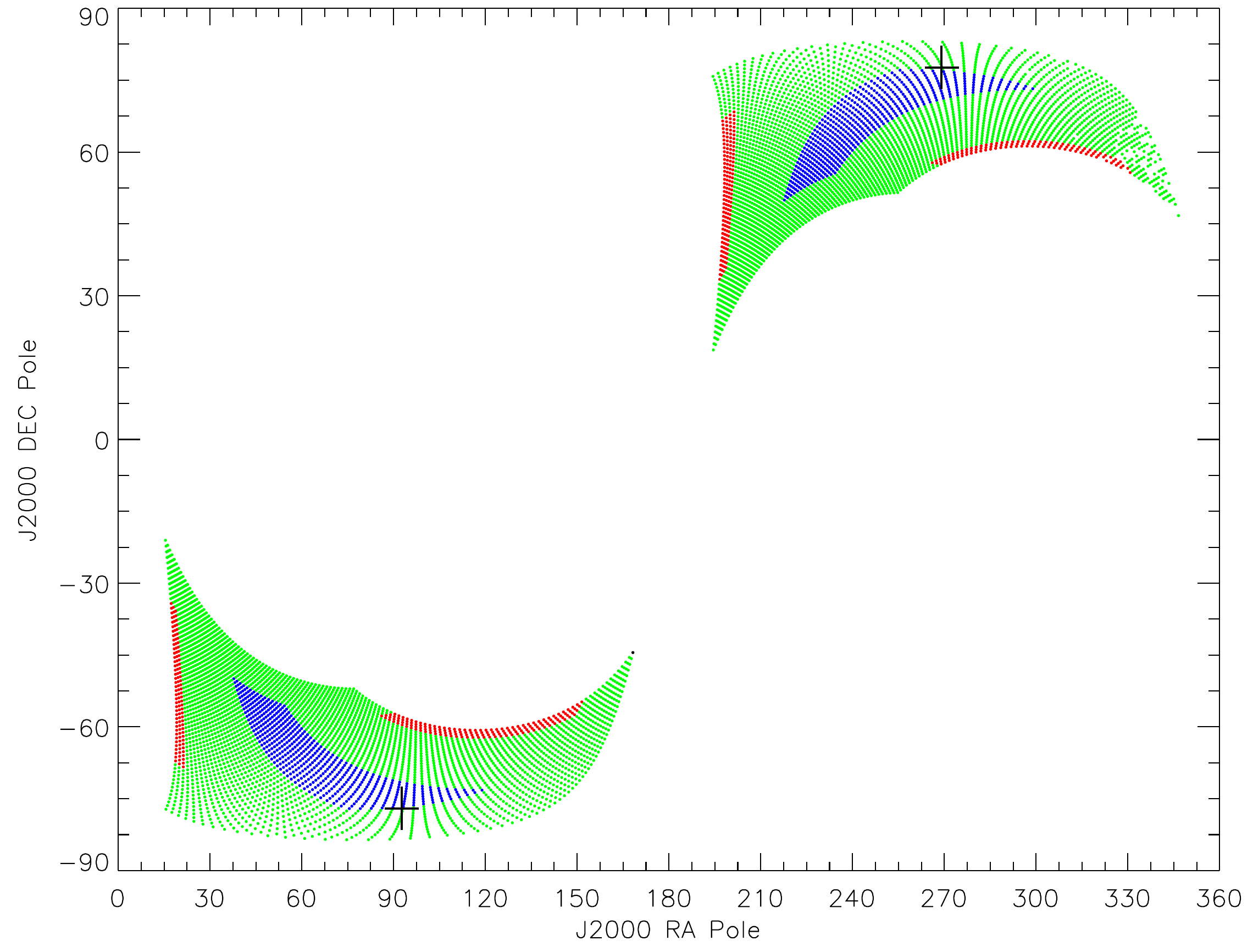}
  \caption{Solution space for the pole of rotation for
     2013/2021 (left) and 2015/2021 (right), see text.
     The black cross gives the solution
     intersection of the red and blue areas.}
  \label{fig:pole}
\end{figure*}

  From the change of shape of the light curves collected in 2013, 2015 and 2021, we determine a
  set of two symmetric pole solutions. We therefore solve the following system which gives
  the position of the rotation pole from simple and relevant assumptions on the latitude of the
  sub-observer point \citep{2007Icar..187..482D}.
  The latitude of the sub-observer point (\bsep) and the North pole
  position angle ($n_p$) are related to the equatorial coordinates of
  the rotation pole ($\alpha_0$, $\delta_0$) and the
  equatorial coordinates of the asteroid ($\alpha$, $\delta$) for each epoch by the following equations :

\begin{eqnarray}
  \sin \bsep &=& -\sin \delta_0 \sin \delta - \cos \delta_0 \cos \delta \cos(\alpha-\alpha_0) \nonumber\\  
  \sin n_p \cos \bsep &=& -\cos \delta_0 \sin( \alpha-\alpha_0) \\
  \cos n_p \cos \bsep &=& \sin \delta_0 \cos \delta - \cos \delta_0 \sin \delta \cos(\alpha-\alpha_0) \nonumber
\end{eqnarray}

  The search for possible solutions for the rotation pole of the asteroid is mainly based on
  assumptions concerning the latitude of the sub-observer point. These are constrained
  by the observed amplitudes of the light curves.
  The light curve observed in 2013 is the one with the lowest amplitude, 0.76 mag. That of
  2015, conversely, presents the largest amplitude, 1.16 mag. Finally, that of 2021 still shows a
  significant amplitude although smaller than in 2015. From these findings we 
  formulate the following assumptions :
  in 2013, $|\bsep| >$ 14\degr ;
  in 2015, $|\bsep| <$  4\degr ;
  in 2021, $|\bsep| <$ 10\degr.

  We consider the two following observation pairs : 2013/2021 and 2015/2021. For each pair, we
  search graphically for the areas of the solution space for which the previous conditions on the
  latitude of the sub-observer point are satisfied. Each pair of subterrestrial latitude values for each
  epoch point is defined on a grid of values from -16\degr~to +16\degr~with a step of
  0.5\degr~(\Autoref{fig:pole}).

  In order for the above conditions to be satisfied,
  it is necessary to select the solutions that are at the
  intersection of the red area for the 2013/2021 epoch
  (\Autoref{fig:pole}, left) and the blue area for the
  2015/2021 epoch
  (\Autoref{fig:pole}, right).
  We can then infer two symmetrical pole solutions (direct and retrograde), they
  are visualized on \Autoref{fig:pole} with a black cross.
  The J2000 equatorial coordinates of the pole 1 are
  $\alpha_0 = 92 \pm 2$\degr~and
  $\delta_0 = -77 \pm 2$\degr.
  The solution for the pole 2 is given by
  $\alpha_0 = 268 \pm 2$\degr~and
  $\delta_0 = +77 \pm 2$\degr. 

\subsection{Shape}

  High brightness variations (greater than 0.9 mag), U-shaped maxima and V-shaped minima are
  convincingly suggestive of an elongated shape with two lobes at the ends separated by a narrower
  neck \citep{2015Icar..245...64D}.
  In a previous study devoted to the Trojan asteroid Thymbraeus \citep{2007AJ....134.1133M},
  the authors sought to model their photometric light curves
  using two tightened and doubly synchronized equilibrium Roche ellipsoids.
  The aim was to determine how well the observations could be matched by theoretical light curves of a
  bilobated shape.

  However, \citet{2010ApJ...719.1602G} showed that equilibrium figures of tightly
  bound binaries are no longer triaxial ellipsoids, and departures from the pure ellipsoidal forms may
  amount to nearly 20\%. They found that at mutual separation on the order of twice the sum of their
  mean radius, departures from ellipsoids given by the Roche binary approximation are negligible.
  On the other hand, \citet{2015Icar..245...64D} showed 
  \revadd{
    two Roche ellipsoids only provided an approximation
    to the properties of a bilobated object,
    while dumbbell-shaped equilibrium figures provide
    numerical solution without bias on the angular momentum.}
  %
  In such case, the solution is entirely described by a single parameter,
  the normalized angular velocity $\Omega$ defined by
  the ratio between the angular frequency $\omega$ 
  and the critical spin rate for a spherical body $\omega_c$, which is the
  maximum spin rate that can be sustained by \revadd{a} rigid body :

\begin{equation}
  \Omega = \frac{\omega}{\omega_c} = \omega \left/ \sqrt{ \frac{4}{3} \pi \rho G} \right.
\end{equation}

  \noindent where $G$ is the gravitational constant, and 
  $\rho$ the bulk density.

\begin{figure}[t]
    \centering
    \input{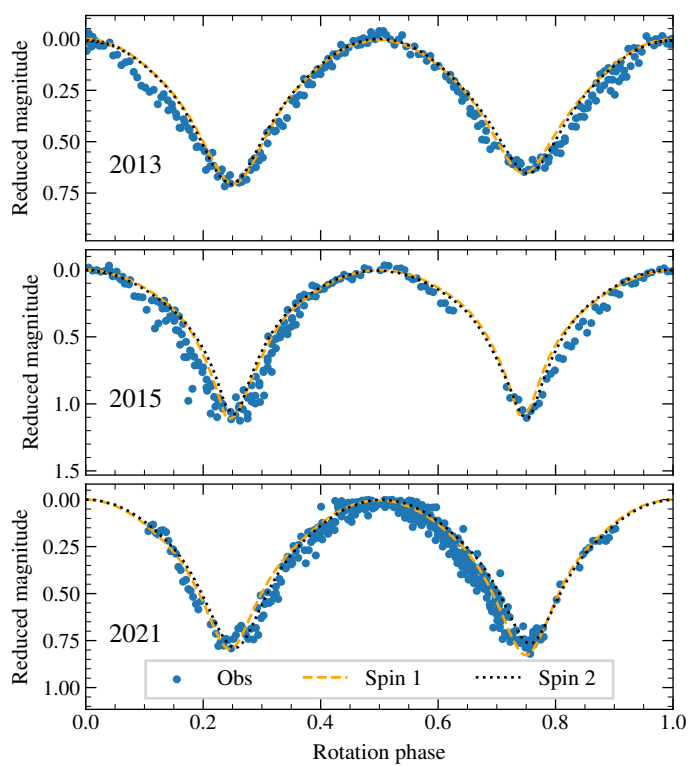}
    \caption{Observed light curves of Thymbraeus compared with the synthetic
      light curves with the two spin solutions. 
      \revadd{The coordinates ($\alpha_0$, $\delta_0$) of the spin solutions are
      (92\degr,-77\degr) and (268\degr,-77\degr) for the Spin 1 and 2, respectively
      (see text).}
      }
    \label{fig:lc}
\end{figure}

  Therefore, we investigate here a more reliable shape solution belonging to the dumbbell equilibrium
  sequence. The objects of this sequence are symmetric with respect to one axis and rotate around a
  second axis perpendicular to the symmetry axis. The dumbbell sequence was first computed by
  \citet{1982PThPh..67.1068E} and more recently fully characterized by \citet{2015Icar..245...64D}.
  The synthetic
  light curves are produced taking into account the photometric effects induced by the scattering
  effects of sunlight by the surface of the object coupled to the phase angle.
  \revadd{We present these lightcurves with the observations in \Autoref{fig:lc}.
  The lightcurves agree well with observations (RMS residuals of
  0.05, 0.12, 0.6\,mag for the three epochs), 
  while presenting some departures, likely due to surface features not represented
  by the dumbbell equilibrium figure.}
  In addition, \revadd{even at the}
  small phase angles involved
  (8\degr~in 2013,
   6\degr~in 2015 and 
   4\degr~in 2021), it is necessary to take into
  account the significant effect of mutual shadowing.
  We adopted a scattering law combining
  through a weight factor $k$ a lambertian icy-type law,
  suitable for high albedo surfaces, and a
  lunar-type reflection described by the Lommel-Seeliger law appropriate for low albedo surfaces
  \citep{2001Icar..153...37K}.
  We adopt $k$ = 0.05. The best-fit solution was obtained simultaneously
  with the determination of the sidereal periods for each pole solution:
  $\Omega$\,=\,0.285$\pm$0.01,
  P\textsub{sid,1}\,=\,12.671821\,h and 
  P\textsub{sid,2}\,=\,12.672607\,h.
  We determine a density of 
  $\rho = \Dens \pm \dDens$\,\sid{}
  from the sidereal periods and the normalized angular velocity.
  This low density is \revadd{similar to the density of $780_{-80}^{+50}$\,\sid{}
  originally reported by \citet{2007AJ....134.1133M}, 
  and} typical of Trojans and similarly-sized 
  KBOs \citep[e.g.,][]{2012PSS...73...98C, 2015-AsteroidsIV-Scheeres} and 
  suggests a porous interior characteristic of rubble piles.

  The light curves present an asymmetry between the minima, noticeably
  apparent in 2013 with a magnitude
  differential of 0.05 mag. We thus apply a small perturbation to the
  hydrostatic equilibrium shape
  solution using a Gaussian random sphere 
  \citep{1998A&A...332.1087M} allowing to take into account
  substantial internal friction present in rubble pile objects
  \citep{2016Icar..265...29D}.
  In doing so, a so-called
  near-equilibrium shape is constructed by combining the initial dumbbell shape model with a
  Gaussian random sphere which approximates the departure with the
  real shape. Obviously, this does not mean that the resulting solution is
  the exact solution but just
  that the asymmetry between photometric minima may be interpreted by small shape deviations from
  a perfect fluid solution. We use a Gaussian random sphere generated by two parameters, the
  relative standard deviation of radial distance, $\sigma$ = 0.05, and the input correlation angle of the
  Gaussian sphere, $\Gamma$= 180\degr. The resulting object has the following statistical properties according to
  the notations introduced in 
  \citet{1998A&A...332.1087M} 
  and
  \citet{1998A&A...333..753M}:
  $\tilde{\sigma}$ = 0.98,
  $\tilde{\Gamma}$ = 56.9\degr,
  $\tilde{\rho}$ = 0.86, and the standard deviation of shape angle, 
  $\tilde{\Phi}$ = 40.8\degr.

\begin{figure}[t]
    \centering
    \includegraphics[width=\hsize]{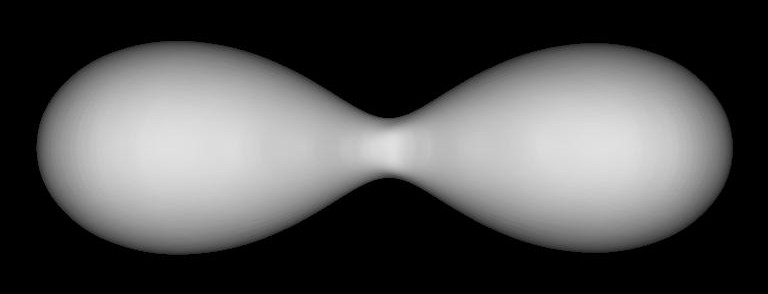}\\
    \includegraphics[width=\hsize]{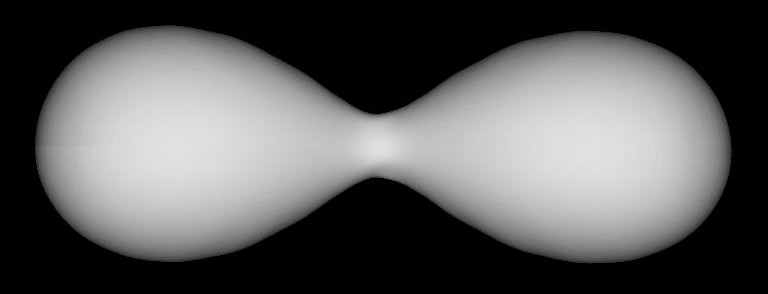}
    \caption{Shape model of Thymbraeus, see from the
    equator (top) and the spin axis (bottom).
    }
    \label{fig:shape}
\end{figure}

  The inferred estimated
  slope angle (assimilated to the angle of repose) is 2.5\degr.
  The angle of repose for a fluid body is, however, strictly zero.
  It is often pointed out 
  that loosely consolidated piles of aggregated particles
  have slopes that are maintained at the angle of repose with respect to horizontal.
  We present in \Autoref{fig:shape} the final shape solution obtained for Thymbraeus.

\begin{figure}[t]
    \centering
    \input{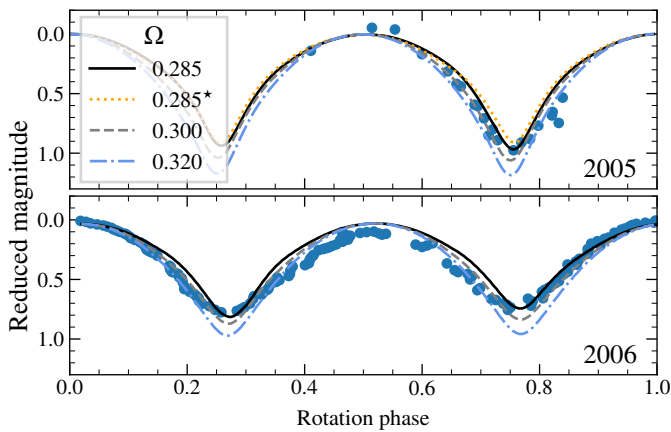}
    \caption{Effect of $\Omega$ on the light curve, compared to the
     observations of 2005 and 2006 from \citet{2007AJ....134.1133M}.
     The solution marked with a star ($\star$) does not account
     for self-shadowing.
    }
    \label{fig:omega}
\end{figure}

  Our model reproduces faithfully the observed light curves
  without invoking two Roche ellipsoids with a significant secondary-to-primary 
  mass ratio as for the solution proposed by \citet{2007AJ....134.1133M}.
  Our solution also reproduces the light curves observed in April 2005 and
  February 2006 and published in \citet{2007AJ....134.1133M},
  see \Autoref{fig:omega}.
  The photometric ranges and the asymmetries
  between the minima are perfectly reproduced.
  \citet{2007AJ....134.1133M} assumed that the object was viewed
  equatorially in 2005 (aspect angle of 90\degr or \bsep = 0\degr),
  thus producing the larger photometric range
  of $\sim$1\,mag but at the cost of a differential drop between 
  minima of nearly 0.1\,mag. With the 2006
  observations, they found that an aspect angle of 75\degr 
  (\bsep=15\degr) produced a better fit. Our solution
  gives respectively $\bsep = -6$\degr~in 2005 and $\bsep = -10$\degr~in 2006.
  Furthermore, if we do not take into
  account the effects of cast shadows, we obtain an amplitude of 0.917\,mag in 2005 instead of
  0.971\,mag and with a quasi-absence of asymmetry 
  (\Autoref{fig:omega}).
  We also plot the synthetic light curves for
  different values of $\Omega$. 
  They show that the photometric range increases with $\Omega$ while the differential
  in magnitude decreases.
  This results from the fact that when $\Omega$ increases, the corresponding
  equilibrium figure of the dumbbell sequence elongates with a thickening of its waist. With the
  nominal solution $\Omega=0.285$,
  the magnitude differential is $\Delta=0.034$\,mag, but for $\Omega=0.300$,
  $\Delta=0.017$\,mag.

  All the collected light curves so far do not show unequal minima, this tends to show that the two
  lobes are of similar size. If this were not the case, we would observe a differential in magnitude
  whatever the orientation of the system. This proves that the magnitude differential arises from a
  significant mutual shadowing between the lobes, differing one from the other by their shape but not
  by their size, under specific geometric configurations. This underlines the importance of taking into
  account all photometric effects including mutual shadowing which must be combined
  simultaneously with a reliable pole solution, independently derived from any consideration on the
  shape model, and a realistic shape solution.

\section{Discussion\label{sec:disc}}

  The dumbbell model had already been successfully applied to the asteroid (216) Kleopatra
  \citep{2015Icar..245...64D},
  which was already known from radar imagery to have two lobes at its extremities
  \citep{2000Sci...288..836O},
  earning it the nickname of ``dog bone''.
  However, the radar model could not satisfactorily
  account for the photometric observations which required to take into account the
  effects of self-shadowing.
  The stellar occultation observations confirmed that the radar model was not sufficiently
  elongated and that its central waist was narrower \citep{2011Icar..211.1022D}.
  More recently, new high resolution imaging made
  with the ESO VLT SPHERE/ZIMPOL camera, confirmed that the shape of (216) Kleopatra is very close
  to an equilibrium dumbbell figure with two lobes and a slightly thicker waist
  \citep{2021A&A...653A..57M}
  and $\Omega$=0.334, 
  only slightly higher than the value of 0.297 found by \citet{2015Icar..245...64D}.
  Consequently, the
  dumbbell equilibrium figure formalism seems to be a trustworthy approach and our dumbbell
  model of Thymbraeus appears to be the best suited to explain the photometric light curves.

  The
  presence of two large lobes, separated by a narrower central part, that are roughly identical in size
  but different in shape is supported by the importance of self-shadowing effects in photometric
  observations, without which it is impossible to account for the difference in magnitude drop
  between the minima of some light curves. Such a physical feature is rare and is a key point to
  understand the origin and future of this striking shape. It now needs to be confirmed with new high
  resolution observations either by precision photometry or by stellar occultations. 
  Precision photometry should also allow to discriminate the two pole solutions.
  
\begin{figure}[t]
    \centering
    \input{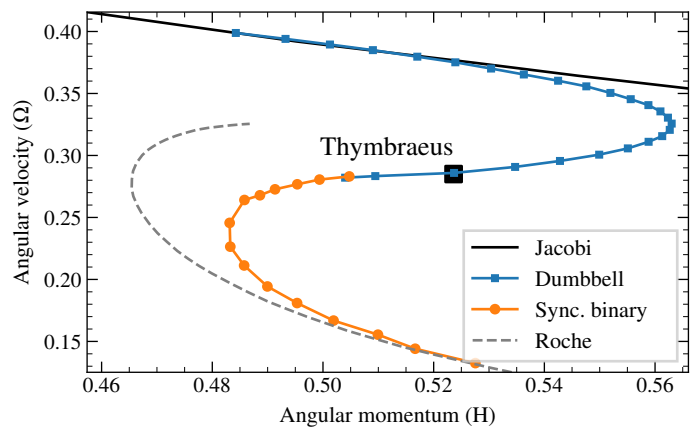}
    \caption{\revadd{Angular velocity $\Omega$
    against angular momentum H for Thymbraeus.
    The solid curves represent the Jacobi, 
    dumbbell \citep{2015Icar..245...64D},
    and twin synchronous binary ellipsoids
    \citep{2010ApJ...719.1602G}
    sequences. The Roche approximation is represented
    by the dashed curve.
    Figure adapted from \citet{2015Icar..245...64D}.
    }}
    \label{fig:sequence}
\end{figure}

  \revadd{The shape solution found in this work is at the end of the dumbbell equilibrium sequence
  (\Autoref{fig:sequence}).
  The angular momentum H is computed as $\frac{2}{5} \lambda \Omega$, with 
  $\lambda$ the non-sphericity parameter \citep{2015Icar..245...64D}, equal to 4.5932 
  for Thymbraeus.
  }
  This sequence ends for the value $\Omega$=0.2815 and joins at this point the sequence
  of synchronous congruent binaries numerically investigated by \citet{2009Icar..200..636S}
  and more completely by \citet{2010ApJ...719.1602G}.
  Furthermore, the bulk density derived from the model is very close to that
  determined for another Trojan asteroid,
  (617) Patroclus \citep{2020Icar..35213990B}, which is a doubly
  synchronous system. \revadd{Should} Thymbraeus rotation be accelerated, 
  it would fission and produce
  a doubly synchronized system.

\section{Conclusions}

  We collected light curves of the Trojan (17365) Thymbraeus in 2015 and
  2021, and retrieve observations from 2005, 2006, and 2013. These
  observations present periodic large-amplitude variations, hinting
  at the binarity nature of Thymbareus.
  We analyze these light curves with the formalism of dumbbell equilibrium
  figures. We determine Thymbareus to be a \revadd{bilobated asteroid}, with
  two lobes of equal size but differing shapes.
  Its sidereal rotation is found to be 12.672\,h,
  and two symmetric poles corresponding
  to the direct and prograde rotation are determined at
  J2000 equatorial coordinates 
  ($\alpha_0$,$\delta_0$) of
  (92\degr, -77\degr) and
  (268\degr, +77\degr), respectively, 
  with an uncertainty of 2\degr.
  The density of Thymbareus is found to be 
  $\Dens \pm \dDens$\,\sid{}, 
  confirming the original report by \citet{2007AJ....134.1133M} and similar to that of
  other Jupiter Trojans and small Kuiper-Belt objects.
  The rotation of Thymbareus is close to the end of the dumbbell
  equilibrium sequence. A faster-rotating Thymbareus would
  fission into an equal-size binary reminiscent of 
  (617) Patroclus.

\begin{acknowledgements}
  We thank the AGORA association which administrates the
  60 cm telescope at Les Makes observatory, La Reunion island, under a financial
  agreement with Paris Observatory. Thanks to A. Peyrot, J.-P. Teng
  for local support, and A. Klotz for helping with the robotizing.\\

  TRAPPIST-South is funded by the Belgian Fund for Scientific Research
  (Fond National de la Recherche Scientifique, FNRS) under the grant 
  PDR T.0120.21. TRAPPIST-North is a
  project funded by the University of Liège, and performed in collaboration with
  Cadi Ayyad University of Marrakesh.
  E. Jehin is a Belgian FNRS Senior Research Associate.\\

  The authors acknowledge the use of the Virtual Observatory tools
  \texttt{Miriade}\,\footnote{\texttt{Miriade}: \href{http://vo.imcce.fr/webservices/miriade/}{http://vo.imcce.fr/webservices/miriade/}}
  \citep{2008LPICo1405.8374B}, 
  \ssodnet\,\footnote{\texttt{SsODNet}: \href{https://ssp.imcce.fr/webservices/ssodnet/}{https://ssp.imcce.fr/webservices/ssodnet/}}
  \citep{2023A&A...671A.151B},
  and
  \texttt{TOPCAT}\,\footnote{\texttt{TOPCAT}:
    \href{http://www.star.bris.ac.uk/~mbt/topcat/}{http://www.star.bris.ac.uk/~mbt/topcat/}} and
  \texttt{STILTS}\,\footnote{\texttt{STILTS}: \href{http://www.star.bris.ac.uk/~mbt/stilts/}{http://www.star.bris.ac.uk/~mbt/stilts/}}
  \citep{2005ASPC..347...29T}.
  Thanks to the developers and maintainers.

\end{acknowledgements}

\bibliographystyle{aa} 
\bibliography{current} 


\end{document}